\documentclass[12pt]{aastex}
\begin{document}

\title{How Accurately Can We Calculate the Depth of
  the Solar Convective Zone?}
\author{John N. Bahcall and Aldo  M. Serenelli}
 \affil{Institute for Advanced Study, Einstein Drive,
Princeton, NJ
  08540}
\and
\author{Marc Pinsonneault}
\affil{Department of Astronomy, Ohio State University, Columbus, OH
  43210}

\begin{abstract}
We evaluate the logarithmic derivative of the depth of the solar
convective zone with respect to the logarithm of the radiative
opacity, $\partial \ln R_{\rm CZ}/\partial \ln \kappa$. We use
this expression to show that the radiative opacity near the base
of the solar convective zone (CZ) must be known to an accuracy of
$\pm 1$\% in order to calculate the CZ depth to the accuracy of
the helioseismological measurement, $R_{\rm CZ} = (0.713 \pm
0.001)R_\odot$. The radiative opacity near the base of the CZ that
is obtained from OPAL tables must be increased by $\sim 21$\% in
the Bahcall-Pinsonneault (2004) solar model if one wants to invoke
opacity errors in order to reconcile recent solar heavy abundance
determinations with the helioseismological measurement of $R_{\rm
CZ}$. We show that the radiative opacity near the base of the
convective zone depends sensitively upon the assumed heavy element
mass fraction, $Z$. The uncertainty in the measured value of $Z$
is currently the limiting factor in our ability to calculate the
depth of the CZ. Different state-of-the-art interpolation schemes
using the existing OPAL tables yield opacity values that differ by
$\sim 4$\% . We describe the finer grid spacings that are
necessary to interpolate the radiative opacity to $\pm 1$\%.
Uncertainties due to the equation of state do not significantly
affect the calculated depth of the convective zone.
\end{abstract}


\section{INTRODUCTION}
\label{sec:intro}
%%FOR citation in text use the \nocite{} command.
The depth of the solar convective zone has been measured by
helioseismological techniques to high accuracy.  In the most
comprehensive study to date, Basu \& Antia (1997)\nocite{basucz} have
investigated the influence of  observational and theoretical
systematic uncertainties as well as measurement errors. Basu and
Antia concluded that the base of the solar convective zone
currently lies at a depth of
\begin{equation}
R_{\rm CZ} ~=~ (0.713 \pm 0.001)R_\odot \, .
\label{eq:rczmeasured}
\end{equation}
The result of Basu and Antia is consistent with the earlier
measurements of Kosovichev \& Fedorova
(1991),\nocite{kosovichevcz} who obtained $R_{\rm CZ} ~=~ (0.713
\pm 0.001)R_\odot$, and Christensen-Dalsgaard, Gough, \& Thompson
(1991),\nocite{jcdCZ} who also obtained $R_{\rm CZ} ~=~ (0.713 \pm
0.003R_\odot$, as well as with the determination of Guzik \& Cox
(1993),\nocite{guzikcz} who found $R_{\rm CZ} ~=~ (0.712 \pm
0.001)R_\odot$.  Basu (1998)\nocite{basu98} also studied the
effect of the assumed value of the solar radius on the inferred
depth of the convective zone and found $R_{\rm CZ} ~=~ (0.7135 \pm
0.0005)R_\odot$. The analyses in these different studies span a
wide range of reference solar models and analysis techniques.

On the basis of the analyses cited above, the measurement of the
depth of the solar convective zone appears robust and precise.

Recently, new precision measurements have been made of the C, N, O,
Ne, and Ar abundances on the surface of the Sun (Allende Prieto,
Lambert, \& Asplund 2001;\nocite{allende01} Allende Prieto, Lambert,
\& Asplund 2002;\nocite{allende02} Asplund et al.
2004;\nocite{asplund04} Asplund et al. 2000;\nocite{asplundetal00}
Asplund 2000)\nocite{asplund00}.  These new abundance determinations
use three-dimensional rather than one-dimensional atmospheric models,
include hydrodynamical effects, and pay particular attention to
uncertainties in atomic data and the observational spectra. The new
abundance estimates, together with the previous best-estimates for
other solar surface abundances~(Grevesse \& Sauval
1998)\nocite{oldcomp}, imply $Z/X = 0.0176$, much less than the
previous value of $Z/X = 0.0229$ (Grevesse \& Sauval
1998)\nocite{oldcomp}.

For a solar model with the recently-determined heavy element to
hydrogen ratio, the calculated depth of the convective zone
is~(Bahcall \& Pinsonneault 2004)\nocite{BP04}
\begin{equation}
R_{\rm CZ}(Z/X = 0.0176) ~=~ 0.726 R_\odot \, ,
\label{eq:rcznewzoverx}
\end{equation}
which is very different from the measured depth of the CZ (see
equation~[\ref{eq:rczmeasured}]). On the other hand, Basu and Antia
(2004)~\nocite{basu04} have shown that the helioseismological
determination of $R_{\rm CZ}$, equation \ref{eq:rczmeasured},  is
not affected if one assumes the correctness of the lower heavy
element abundances ($Z/X = 0.0176$).

Something is wrong. We have a new solar problem: ``the convective
zone (CZ) problem.''

The radiative opacity is a key ingredient in calculating the depth
of the convective zone. Moreover, about 95\% of the total
radiative opacity near the base of the convective zone involves
bound electrons, either bound-free or bound-bound opacity
(Iglesias 2004).\nocite{iglesias04} Thus opacity calculations in this region involve
details of the ionization balance and other delicate atomic
physics properties.

 In this paper, we focus on determining the accuracy with
which the opacity near the base of the convective zone must be
known in order to calculate precisely the depth of the CZ with a
stellar evolution code.  We also evaluate the accuracy with which
the opacity near the base of the CZ can be interpolated from OPAL
tables. For a related comparison of the Los Alamos LEDCOP
opacities and the OPAL opacities, see Neuforge-Verheecke et al.
(2001)~\nocite{losalamos}. For comprehensive discussions of
stellar radiative opacities, the reader is referred to the
important reviews by Rogers and Iglesias (1998)\nocite{OPALreview}
and by Seaton et al. (1995)\nocite{opbook}.

We investigate in a paper in preparation (Bahcall, Basu,
Pinsonneault, and Serenelli 2004)~\nocite{inpreparation} the
helioseismological implications of the changes in opacity that are
discussed in the present paper. The viability of any proposed
change in the opacity discussed in the present paper must be
tested by comparing a solar model that is evolved with the changed
opacity with a complete set of precise helioseismological data.
There is no compelling reason to believe that the illustrative
change in opacity considered here, which is highly peaked in
radius,  will be either reproduced exactly by new opacity
calculations or will be precisely consistent with
helioseismological constraints. In the future, once new opacity
calculations are available that satisfy the requirements described
in this paper, it will be possible to test simultaneously the new
opacities, the solar model evolution, and the helioseismological
implications.

We derive in \S~\ref{sec:dependence} the dependence, $\partial \ln
R_{\rm CZ}/\partial \ln \kappa$, of the calculated depth of the
solar convective zone upon the assumed radiative opacity. We apply
this result to determine the accuracy with which the opacity must
be known in order to calculate the depth of the CZ to the accuracy
with which it is measured helioseismologically.  We also determine
the change in the standard OPAL opacity that is required in order
to reconcile the helioseismological measurement with the recent
determinations of heavy element abundances. We evaluate in
\S~\ref{sec:composition} the dependence of the radiative opacity
near the base of the convective zone upon the stellar composition.
We find that the opacity depends sensitively upon the assumed
heavy element abundance. We compare in \S~\ref{sec:comparison} the
opacities obtained from two different interpolation schemes that
are both applied to the same published OPAL opacity tables.
Throughout this paper, we adopt the OPAL opacities (Iglesias \&
Rogers 1991a,b; Rogers \& Iglesias 1992; Iglesias \& Rogers
1996)\nocite{iglesias91a,iglesias91b,rogers92,opalopacity96} as
standard, when supplemented at low temperatures by values from
Alexander \& Fergusson (1994)\nocite{alexanderopac}. We use
simulated opacity tables in \S~\ref{sec:shifted} to estimate the
likely uncertainties that result from interpolations within the
existing OPAL opacity tables and to determine the grid sizes to
obtain small interpolation errors. For completeness and for
contrast, we use four different equations of state  to show in
Appendix~A that uncertainties due to the choice of EOS are not
important, at the present level of accuracy, for the calculation
of the depth of the solar convective zone. We also demonstrate in
Appendix~B that uncertainties in the nuclear reaction rates affect
the depth of the solar convective zone only at the level of 0.1\%
. In Appendix~C, we verify that the conversion of carbon and
oxygen in CNO burning, which cannot be accurately modeled with
existing opacity codes, causes a 0.1\%  uncertainty in the
calculated depth of the convective zone. Basu and Antia (2004) (see
also Asplund et al. 2004)  have shown that errors in the
calculation of the diffusion coefficients are unlikely to be the
correct explanation of the discrepancy between measured and
calculated depth of the solar convective zone. Other solar model
ingredients, including the element diffusion coefficients, can
affect the calculated depth of the convective zone. A complete
investigation of all the possible effects on the convective zone
is beyond the scope of the present paper and would distract the
reader from our main concern, the effect of the radiative opacity.
Moreover, we believe that the radiative opacity and the heavy
element abundance provide the single largest contributions to the
error budget for the calculation of the solar convective zone. The
effect of the heavy element abundance on the calculated depth of
the convective zone has been evaluated in Bahcall and Pinsonneault
(2004). We summarize and discuss our main results \hbox{in
\S~\ref{sec:summary}}.

\section{DEPENDENCE OF CALCULATED DEPTH OF CONVECTIVE ZONE ON RADIATIVE
OPACITY} \label{sec:dependence}

In this section, we determine the dependence of the calculated
depth of the solar convective zone upon the value of the radiative
opacity in the vicinity of the base of the convective zone. Using
this dependence, we then answer two questions. First, how
accurately must the radiative opacity be known in order to
calculate the depth of the convective zone to the accuracy with
which the depth is measured by helioseismology? Second, how large
must the error in the radiative opacity at the base of the solar
convective zone be in order to explain the difference between the
measured value of $R_{\rm CZ}$ and the value of $R_{\rm CZ}$ that
is calculated in a solar model that is constructed using the
recently determined heavy element abundances ($Z/X = 0.0176$)?

We follow the approach introduced by Bahcall, Bahcall, \& Ulrich
(1969)\nocite{sensitivity} in  which we change the standard (OPAL)
opacity in the vicinity of the CZ by a small functional amount
that depends upon an adjustable parameter $\alpha$.  We calculate
a series of solar models for different values of $\alpha$, which
permits us to evaluate the logarithmic derivative of $R_{\rm CZ}$
with respect to the opacity near the base of the CZ. We begin with
a brief description of the solar models used in our studies.

\subsection{Description of the solar models}
\label{subsec:solarmodel}

% To here

The solar age adopted in this article  is $4.57 \times 10^9$ yr.
At this age, the solar models are required to have the present
values for the solar luminosity (L$_\odot$), the solar radius
(R$_\odot$), and the ratio of heavy elements to hydrogen by
mass($Z/X$). We adopt the values L$_\odot= 3.8418 \times 10^{33}$
ergs$^{-1}$, $R_\odot= 6.9598 \times 10^{10}$ cm,  and $Z/X=
0.0229$ respectively (see Bahcall, Pinsonneault,  \& Basu 2001).
The models are calculated using the OPAL equation  of state
(hereafter OPAL 1996; Rogers, Swenson, \&  Iglesias
1996\nocite{opalEOS96}) unless stated otherwise, and the  OPAL
opacities (see \S~\ref{sec:intro}). The nuclear reaction rates are
those used  in Bahcall et al. (2001). Element diffusion is
incorporated  for helium and  metals (Thoul, Bahcall, \& Loeb
1994).\nocite{thoul}  We use the mixing length theory  for convection and  the
Schwarzschild criterion to determine the location of the
convective boundaries.

\begin{table} [!t]
\begin{center}
\caption{The adopted compositions used for the computation of
solar models BP04+ and BP04 (and variations thereupon). The
relative abundances given in the table denote Log~$N_i$ in the
usual scale in which Log $N_H$= 12. We use meteoritic abundances
when available. In the past, when conflicts between meteoritic and
atmospheric abundances have existed, the meteoritic determinations
have often turned out to be more correct.
 \label{tab:composition}}
\vglue.2in
\begin{tabular}{lcclcc}
\hline\hline
Elem. & BP04+ & BP04 & Elem. & BP04+ & BP04\\
\hline
C & 8.39 & 8.52 & Cl & 5.28 & 5.28\\
N & 7.80 & 7.92 & Ar & 6.18 & 6.40\\
O & 8.69 & 8.83 & K & 5.13 & 5.13 \\
Ne & 7.84 & 8.08 & Ca & 6.35 & 6.35 \\
Na & 6.32 & 6.32 & Ti & 4.94 & 4.94 \\
Mg & 7.58 & 7.58 & Cr & 5.69 & 5.69 \\
Al & 6.49 & 6.49 & Mg & 5.53 & 5.53 \\
Si & 7.56 & 7.56 & Fe & 7.50 & 7.50 \\
P & 5.56 & 5.56 & Ni & 6.25 & 6.25\\
S & 7.20 & 7.20 \\
\hline
\end{tabular}
\end{center}
\end{table}

The adopted heavy element composition is, as discussed in Bahcall
and Pinsonneault (2004), one of the most important ingredients in
determining the value of the solar convective zone that is
obtained from a stellar evolution code. For specificity, we show
in Table~\ref{tab:composition} the specific composition that has
been adopted in computing the models referred to as solar model
BP04+ (see Bahcall and Pinsonneault 2004; includes recent
composition determinations described in: Allende Prieto, Lambert,
\& Asplund 2001;\nocite{allende01} Allende Prieto, Lambert, \&
Asplund 2002;\nocite{allende02} Asplund et al.
2004;\nocite{asplund04} Asplund et al. 2000;\nocite{asplundetal00}
Asplund 2000)\nocite{asplund00} and solar model BP04 (see
Bahcall and Pinsonneault 2001; composition from Grevesse and
Sauval 1998). OPAL opacities have been computed for the same
compositions. The atomic masses on the OPAL website were used in
conjunction with these abundances to compute Z/X. Although the
most precise details of the composition are not important for the
general issues discussed in this paper,
Table~\ref{tab:composition} permits us to be make clear exactly
what compositions were used in the calculations described in this
paper. This may be helpful to the reader since new composition
determinations are currently appearing at frequent intervals.

For the BP04+ solar model, the initial (final) mass fractions are:
$X_0 =.71564~ (0.74862)$, $Y_0 = 0.26960~ (0.23817)$, and $Z_0 =
0.01476~ (0.01321)$. For the BP04 model, the corresponding mass
fractions are: $X_0 =.70775~ (0.72465)$, $Y_0 = 0.27344~ (0.24335)$,
and $Z_0 = 0.01881~ (0.01692)$. Note that the helium abundances in
the two models are the same to within about 2\%, although the
heavy element mass fractions differ by more than 25\%.

One of the main goals of this paper is to compare the numerical
results obtained for different solar evolution codes. To this end,
we compare the results obtained with the Bahcall-Pinsonneault/Yale
code (see Bahcall and Pinsonneault 2004 and references contained
therein) with the Garching/Weiss stellar evolution code. For
further details about the Garching stellar evolution code, we
refer the reader to Schlattl, Weiss, \& Ludwig
(1997)\nocite{garching}, Schlattl (2002)\nocite{schlattl02}, and
references therein.

\subsection{Evaluation of ${\partial \ln R_{\rm CZ}}/{\partial
\ln \kappa}$} \label{subsec:partialderivative}

 For relatively small changes in the radiative opacity, the
 sensitivity to opacity of the calculated depth of the solar
 convective zone can be expressed in terms of a single numerical parameter
  $\beta$, which is defined by the relation
\begin{equation}
\beta~\equiv~ \frac{\partial \ln R_{\rm CZ}}{\partial \ln
\kappa}\, . \label{eq:betadefn}
\end{equation}
To evaluate $\beta$, we multiply the OPAL opacity in the vicinity
of the convective envelope boundary by a Lorentzian function
$f(T)$ given by
\begin{equation}
f(T)= 1 + \frac{\alpha \gamma^2}{\left( (T-T_0)^2 + \gamma^2
\right)} \, .
\label{eq:defnofperturbation}
\end{equation}
Here $T$ is the temperature in the solar model. We label each
radial point in the solar model by its corresponding value of $T$.
We can then write for the opacity that $\kappa= \kappa_0 f(T)$,
where $\kappa_0$ is the unperturbed radiative opacity, $\alpha$ is
the amplitude of the perturbation, and $\gamma$ is the width of
the perturbation (defined as the point where the perturbation
drops to $\alpha /2$).

At the present solar age, the temperature at the base of the CZ is
$T\approx 2.18\times 10^6$K, so this value is adopted for $T_0$.
We calculated solar models for $\gamma= 0.2\times 10^6 {\rm \, K}
\approx 0.1T_0$ and $\alpha= 0, \ \pm 0.030, \ \pm0.060$, which
were well represented by a fixed value of $\beta$.

We find

\begin{equation}
\beta ~=~ -0.095 ~ =~\frac{\partial \ln R_{\rm CZ}}{\partial \ln
\kappa}\, , \label{eq:betavalue}
\end{equation}
or, equivalently,
\begin{equation}
\left( \frac{R_{\rm CZ}}{R_{\rm CZ,0}} \right) =\left(
\frac{\kappa}{\kappa_0} \right)^{-0.095}. \label{eq:powerlaw}
\end{equation}
Since we have used converged solar models that satisfy the
observational constraints on the luminosity, the chemical
composition, and other parameters, the result given in
equation~(\ref{eq:powerlaw})includes all of the feedback effects
required by the boundary conditions and the external observational
constraints.

To test the robustness of this result, we doubled the value of
$\gamma$ to $\gamma= 0.4\times 10^6 {\rm \, K}$ and obtained for
this broader perturbation a similar value for $\beta$, namely,
$\beta = -0.10$ (instead of $-0.095$).  Of course, we do not know
{\it a priori} the exact form of any future change in the
radiative opacity that may result from further research.
Nevertheless, we can conclude from the examples we have studied
that equation~(\ref{eq:powerlaw}) is a good approximation to
changes in the opacity that are local and peaked at the base of
the convection zone.

The sign of $\beta$, which is given in
equation~(\ref{eq:betavalue}), is evident from physical reasoning.
The magnitude of the radiative temperature gradient is
proportional to the opacity since the radiative flux passing
through a given point in the star is fixed. If the radiative
opacity is increased at a fixed point, then the radiative gradient
is increased and the condition for convective stability becomes
more difficult to satisfy. The star can become convectively
unstable at a smaller radius (higher temperature).

The changes in opacity considered here will necessarily bring
about small changes in the inferred surface mass fraction of
hydrogen. Quantitatively, we find analogous to
equation~(\ref{eq:betavalue}) that $(X/X_0)=
(\kappa/\kappa_0)^{-0.023}$, i.e. a  weak dependence. A change of
20\% in opacity leads to an estimated change in X of about 0.4\%,
less than the uncertainties  in the helioseismological
determinations of $X$.

\subsection{How Accurately Do We Need To Know the Opacity?}
\label{subsec:howaccurately}

Equation~(\ref{eq:betavalue}) and equation~(\ref{eq:powerlaw}) imply
that in order to calculate the depth of the convective zone to the
accuracy with which the depth is measured, 1 part in 713, one must
know the radiative opacity at the base of the CZ to an accuracy

\begin{equation}
\left(\frac{\Delta \kappa}{\kappa}\right)_{\rm equivalent~
experimental~accuracy} ~=~ 1\% \, . \label{eq:opacityaccuracy}
\end{equation}
This is extremely high precision for a calculated radiative
opacity, probably beyond the reach of existing calculations.

If we try to explain the difference between the measured value of
$R_{\rm CZ}$ (see equation~[\ref{eq:rczmeasured}]) and the value
calculated using recently determined heavy element abundances (see
equation~[\ref{eq:rcznewzoverx}]), then we find that the opacities
used in the solar model must be in error by
\begin{equation}
\left(\frac{\Delta \kappa}{\kappa}\right)_{\rm Z/X~=~0.0176} ~=~
21\% \, . \label{eq:errornewzoverx}
\end{equation}
The result shown in equation~(\ref{eq:errornewzoverx})applies for
the Bahcall and Pinsonneault (2004) solar model BP04. The Garching
code leads to a slightly larger discrepancy between calculated and
measured depth of the convective zone (cf.
\S~\ref{sec:comparison}).

We have evolved a solar model based upon the recent abundance
determinations, BP04+, but with a 21\% increased opacity near the
base of the convective zone.  The calculated depth of the
convective zone is $R_{\rm CZ} = 0.7133 R_\odot$, in good
agreement (by design) with the measured value. The initial (final,
surface) mass fractions for this model are: $X_0 =.71621~
(0.74776)$, $Y_0 = 0.26919~ (0.23908)$, and $Z_0 = 0.01460~
(0.01316)$. The current surface abundance of Y implied by this
model is about $3\sigma$ smaller than the value determined by Basu
and Antia (2004) from helioseismology. We are not sure how to
regard this discrepancy since the overwhelmingly dominant error in
the helioseismological value is systematic, not statistical.  In
the forthcoming paper by Bahcall et al. (2004), we will compare
the BP04+ solar model with increased opacity with all of the
available helioseismological data.

The estimate given in equation~(\ref{eq:errornewzoverx}) is based
upon the assumption that the opacity is changed only locally,
i.e., near the base of the convective zone (see
equation~[\ref{eq:defnofperturbation}]). If, instead, one changes
the opacity by changing the surface heavy element abundance,
$Z/X$, then the opacity is affected throughout the solar model and
the change required near the base of the convective zone can be
different. The inputs to the models BP04 and BP04+ of Bahcall \&
Pinsonneault (2004) differ only in the assumed heavy element
abundance. BP04+ was calculated assuming $Z/X = 0.0176$ (recently
determined low heavy element abundance) and BP04 was calculated
using $Z/X = 0.0229$ (Grevesse \& Sauval 1998).  Using the results
obtained from these two models (comparing the calculated
difference in the CZ depth between the two converged solar models
with the difference in radiative opacity at same $T$ and $\rho$
near the base of the CZ), we estimate that the opacity near the
base of the CZ must change by $\simeq 14$\% if the pattern of
opacity changes is similar to that induced by composition changes.

We evaluate in the next section the sensitive dependence of the
radiative opacity on the assumed surface heavy element abundance.

\section{Dependence of Radiative Opacity on Composition}
\label{sec:composition}

In this section, we estimate the dependence of the radiative
opacity near the base of the convective zone upon the stellar
composition. We approximate the opacity as a function of the
hydrogen mass fraction, $X$, and the heavy element mass fraction,
$Z$. Thus $\kappa = \kappa(X,Z)$.

The fractional uncertainty in the opacity may then be written in
the form

\begin{equation}
\frac{d\kappa}{\kappa}~=~\left(\frac{\partial \ln \kappa}{\partial
\ln Z}\right)_X\frac{dZ}{Z}~+~\left(\frac{\partial \ln
\kappa}{\partial \ln X}\right)_Z\frac{dX}{X} \, .
\label{eq:kappaoncomposition}
\end{equation}
We have used the existing OPAL opacity tables to evaluate
numerically the fractional derivatives that appear in
equation~(\ref{eq:kappaoncomposition}). We find

\begin{equation}
\left(\frac{\partial \ln \kappa}{\partial \ln Z}\right)_X
~\approxeq~0.70 \, ; ~~\left(\frac{\partial \ln \kappa}{\partial
\ln X}\right)_Z ~\approxeq~0.15 \, .
\label{eq:compositionderivatives}
\end{equation}
The numerical values for the logarithmic derivatives given in
equation~\ref{eq:compositionderivatives} were determined for conditions
similar to those at the base of the current solar convective zone;
we used  $\log T = 6.34$, $\log \rho = -0.7$, $X = 0.74$, and $Z =
0.0169$. Changing the values of the physical variables at which
the derivatives are evaluated causes only small changes in the
derivatives as long as the changes are restricted to stellar
positions close to the base of the convective zone.

The uncertainty in the opacity is dominated by the uncertainty in
the heavy element abundance, $Z$. If we want to know the opacity
to $1$\%, the accuracy required to calculate the depth of the
solar convective zone to the precision with which the depth is
measured (see equation~[\ref{eq:opacityaccuracy}]), then we have
to determine the surface heavy element abundance to a precision of
1\%. This seems like, at present, an impossibly ambitious demand,
at least for the foreseeable future. The current $1\sigma$
uncertainty in $Z$ is about 15\% (see Bahcall \& Pinsonneault
2004).\nocite{BP04}

In the next two sections, we estimate how accurately the radiative
opacity can be interpolated from the existing OPAL opacity tables.

\section{COMPARISON OF THE RADIATIVE OPACITY OBTAINED FROM TWO
DIFFERENT INTERPOLATION SCHEMES}
\label{sec:comparison}

We compare in this section the radiative opacity values
interpolated from standard OPAL opacity tables by two different
interpolation schemes embedded in two extensively used
state-of-the-art stellar evolution codes. In particular, we
interpolate within the OPAL tables using a 4-point Lagrangian
scheme that is implemented in the Yale/BP stellar evolution code
(Guenther, Jaffe, \& Demarque 1989;\nocite{guenther1989} Pinsonneault, Kawaler,
Sofia, \& Demarque 1989;\nocite{pinsonneault89}
Bahcall \& Pinsonneault 1992, 1995,
2001)\nocite{BP92,BP95,BP00} and a bi-rational spline scheme (Sp\"ath
1995\nocite{spath95}) that   is   implemented   in    the Garching
code   (Schlattl   \&   Weiss 1997\nocite{schlattl97}).

The implementations of these two different interpolation schemes
have been extensively tested. There is no absolute way to evaluate
their accuracy since the precision that is obtained depends upon
the behavior of the function being interpolated.  The two
interpolation schemes have different advantages and disadvantages.

\begin{figure}[!t]
\begin{center}
\includegraphics[scale=.60,angle=90]{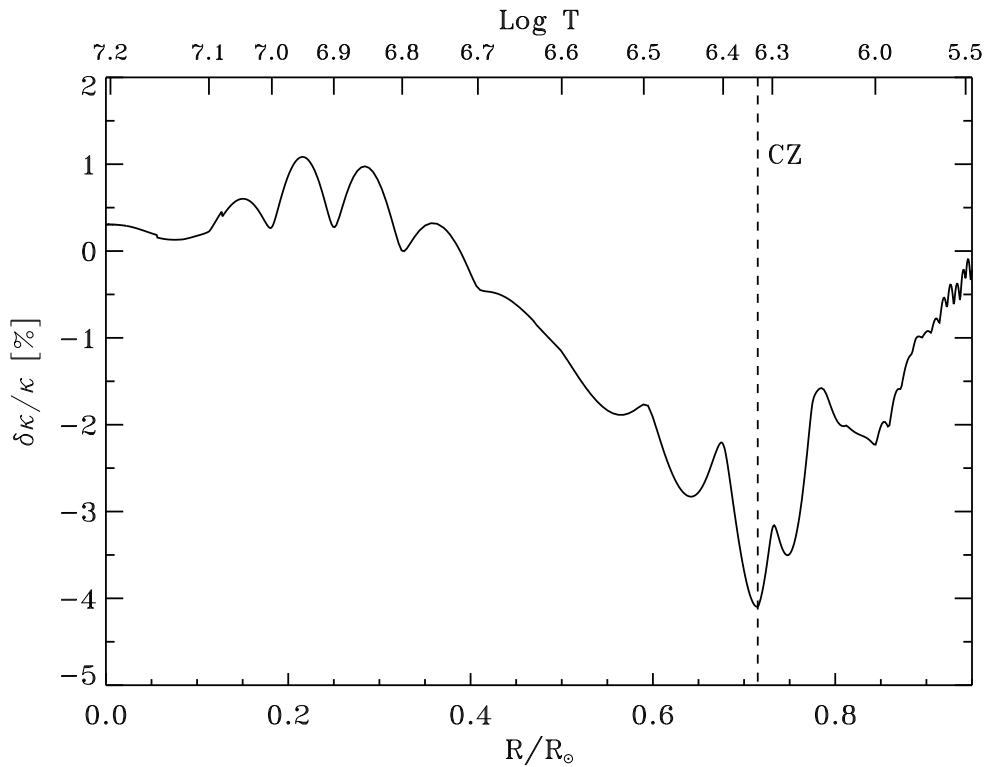}
\end{center}
\caption{Fractional Opacity Difference from Two Interpolation
Schemes. The figure shows the fractional difference in the
opacity, $\delta \kappa /\kappa$ (in percent) obtained from two
interpolation schemes embodied in two widely used stellar
  evolution programs, the Yale/BP code and the MPA code. The
  fractional difference is defined to be $\delta \kappa /\kappa = $
  (Result from bi-rational
  spline $-$ Result from 4-pt-Lagrangian) /
  4-pt-Lagrangian. The figure was made for a fixed T, $\rho$, X, Z profile so
the differences that are shown are only due to
interpolation.}\label{fig:opacitycomparison}
\end{figure}

Figure~\ref{fig:opacitycomparison}  shows the  fractional
difference, $\delta \kappa/\kappa$,  between the  radiative
opacity that  is  obtained using  the Yale/BP  4-point Lagrangian
scheme  and the  opacity  interpolated using  the Garching
bi-rational  spline  scheme  (the damping  parameter  for the
bi-rational splines  was set  to 5). The  fractional opacity is
displayed at different radii  (lower horizontal scale)  and at
different  temperatures (top horizontal scale) in a standard solar
model.  The figure was made for a fixed T, $\rho$, X, Z profile so
the differences that are shown are only due to interpolation.

The amplitude of the difference becomes as large as 4\% near the
base of the convective zone, which is denoted by a vertical line
in Figure~\ref{fig:opacitycomparison}. The interpolated value of
the opacity near the base of the CZ is particularly sensitive to
the interpolation scheme because the temperature of the solar CZ
and the value of $r= \rho / T^3_6$ fall about half way between two
points at which the OPAL opacity is tabulated.

The amplitude of the discrepancy between the two interpolation
schemes is much larger than is permitted if one wants to calculate
the depth of the CZ to the measured accuracy (see
equation~[\ref{eq:opacityaccuracy}]). The above discussion shows that
uncertainties due to interpolation contribute importantly to the
error budget for the calculation of the solar convective zone (see
equation~[\ref{eq:errornewzoverx}]).

\section{SIMULATED OPACITY TABLES: COMPARISON OF INTERPOLATED VALUES
WITH STANDARD VALUES} \label{sec:shifted}

\begin{figure}[!ht]
\begin{center}
\includegraphics[scale=.50,angle=90]{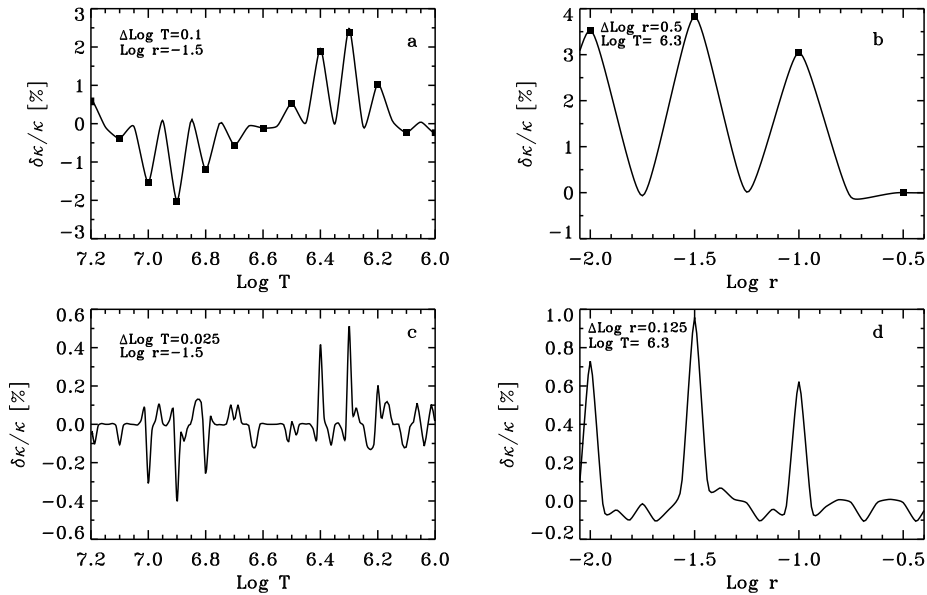}
\end{center}
\caption{Errors in interpolating using shifted opacity. The figure
shows the relative opacity differences, $\delta \kappa / \kappa =
(\kappa_{\rm
  std}- \kappa_{\rm shifted} ) / \kappa_{\rm standard}$, that were
  found between the
interpolated  opacities that were obtained using  the {\it
shifted} and the {\it standard} opacity tables  (see text for
explanation) as a function of the interpolation  variable.  Panel
{\bf a} uses the original OPAL grid spacing  and compares  the
results with the values obtained from shifting the tables in
temperature ($\Delta \log T$). Panel {\bf b} is analogous to panel
{\bf a}, but the opacity tables are shifted in $r$ (defined as
$r=\rho / T_6^3$, shifted in $\Delta \log r$). The squares denote
the location of the tabulated values in the original OPAL opacity
tables. Panels {\bf c} and {\bf d} are similar to panels {\bf a}
and {\bf b}, respectively, but with grid spacings reduced  by  a
factor  of four. Although the regions for $\log T < 6.3$ ($\log r
> -1.5$) are inside the convective envelope, they are shown for
the sake of clarity. In each case, the  grid spacing is  given,
together  with the fixed value  for the  other variable. In  all
cases  the hydrogen and metal mass fractions are  fixed at $X=0.7$
and $Z=0.02$ respectively.}\label{fig:opacityshift}
\end{figure}

In this section, we make plausible estimates of the uncertainties
inherent in interpolating the radiative opacity from the available
OPAL opacity tables. We use simulated opacity tables to make
self-consistency tests of the accuracy of the interpolation
schemes we use.  For specificity, we cite the results obtained
using the bi-rational spline.  Similar results were found with the
4-point Lagrangian spline. The inferences obtained in this section
regarding the accuracy of interpolation within opacity tables
complement and supplement the results obtained in
\S~\ref{sec:comparison} by comparing the outputs of two
different interpolation schemes.

The figures that we show are based upon the following strategy.
Using the existing OPAL opacity tables, we interpolate the value
of the opacity at shifted points, making in this way new but
simulated tables. We then use the simulated tables to predict the
values of the opacity at the original, unshifted points. We take
as one measure of the likely uncertainty the difference between
the opacity values in the original published tables and the
opacity values at the same points obtained by interpolating in the
simulated tables. In all  cases, the largest errors are expected
(and observed) at points originally tabulated in OPAL tables
because none  of the shifted tables (independent of the grid
spacing) have these points tabulated. Since the originally
tabulated values lay in the middle of the shifted intervals, they
give the largest errors.

We have also tested the accuracy of the interpolation schemes by
artificially making the OPAL tables more sparse, i.e., by
omitting points. We then interpolate in the sparser tables to see
how well the interpolation reproduces the omitted values. The
uncertainty estimates obtained using sparser tables are in good
agreement with the uncertainties obtained by shifting points. We
concentrate our discussion here on the results found with the
shifted tables because these results are more easily displayed.

Figure~\ref{fig:opacityshift} shows the fractional differences in
the opacity, $\delta \kappa / \kappa$, that were found between the
values given in the original OPAL tables and the values that were
obtained from the simulated shifted tables. The upper two panels in
the figure use the actual grid spacing of the OPAL tables. The OPAL
tables
are presented in terms of the
logarithm of the temperature ($\log T$) and  the logarithm of $r
\equiv \rho/T_6^3$ ($\log r$).

Figure~\ref{fig:opacityshift}a and Figure~\ref{fig:opacityshift}b
show that the  amplitudes of $\delta \kappa / \kappa$ can be as large as
$3$\% in interpolating within the shifted OPAL tables.

How dense does the opacity grid have to be in order that the
interpolation uncertainty within the grid be less than 1\% (see
equation~[\ref{eq:opacityaccuracy}]) for opacities near the base of
the CZ? In order to provide a plausible answer to this question,
we have  interpolated within simulated opacity tables with grids
of a variety of different densities.

The lower two panels, Figure~\ref{fig:opacityshift}c and
Figure~\ref{fig:opacityshift}d,
 show the results of interpolations within
simulated tables that have grid sizes, respectively, of $\Delta
\log T = 0.025$ and $\Delta r = 0.125$. These simulated tables are
four times as dense in each variable as the existing OPAL opacity
tables. The errors in predicting the unshifted opacity values in
the original OPAL tables are less than 0.6\% (throughout the
physically relevant region) when the values in the simulated
shifted tables are used. With the simple assumptions we have made,
the estimated errors scale approximately linearly with the grid
spacing. However, this linear dependence  results in large part
from our assumption that the opacity values are smooth in $\log T$
and $\log r$ in the regions of interest.

The situation is somewhat different for the heavy element
abundance, $Z$. The existing OPAL opacity tables present values
for three heavy element abundances relevant to the Sun:
 $ Z = 0.01, 0.02, 0.03$. However,
recent redeterminations of the heavy element abundance in the Sun
have suggested that $Z$ is significantly lower than was previously
believed (Allende Prieto, Lambert, \& Asplund 2001,
2002;\nocite{allende01,allende02} Asplund et al.
2004;\nocite{asplund04} Asplund et al. 2000;\nocite{asplundetal00}
Asplund 2000)\nocite{asplund00}. We have experimented numerically
with interpolating in $Z$ within the existing OPAL opacity tables
and also within simulated opacity tables with a denser grid in
$Z$.  We find that the required accuracy (better than 1\%) in
interpolation can be achieved if a grid with $\Delta Z = 0.0025$
is used for values of $Z$ ranging from $Z = 0.0100$ to $Z =
0.0225$. This amounts, in addition to a denser grid, to a shift to
lower values in the mean value of $Z$ that is tabulated.
Fortunately, we find that the existing OPAL grid in the hydrogen
abundance, $X$, is sufficient to permit interpolation in the
opacity to the required accuracy.

\begin{figure}[!ht]
 \begin{center}
 \includegraphics[scale=.60,angle=90]{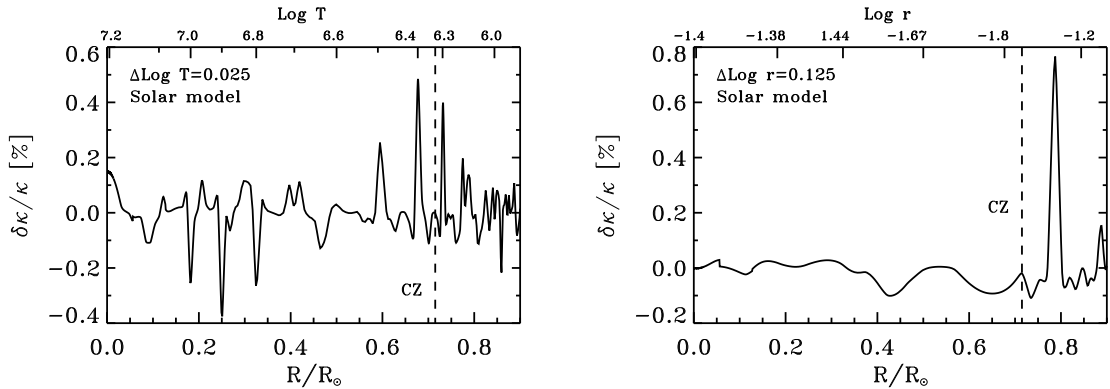}
 \end{center}
 \caption{Interpolation uncertainties in a solar model. The figure
 shows the estimated fractional uncertainties, $\delta \kappa/\kappa$,
 in interpolating
 for a solar model the radiative opacities
 in simulated opacity tables with a grid size $\Delta \log T = 0.025$,
 $\Delta \log r = 0.125$,  and with
 heavy element composition values ranging from $Z = 0.0100$ to
 $Z = 0.0225$ with $\Delta Z = 0.0025$. In Figure~\ref{fig:solarmodel},
the left panel has  $\Delta  \log T=  0.025$ and $\Delta \log r=
0.5$ (original spacing) and the right panel  has $\Delta \log T=
0.1$ (original spacing) and $\Delta \log r = 0.125$.
 The opacities obtained by interpolating in shifted simulated tables
 in $\Delta \log T$, $\Delta \log r$, and Z are
 compared with the values obtained in unshifted simulated tables.
 The differences $\delta \kappa/\kappa$ are shown
 as a function of the fractional solar radius, $R/R_\odot$.
 The upper horizontal axis shows
 the corresponding values of $\log T$ ($\log r$)for the left panel (right panel).
 The location of the base of the CZ
 zone is shown by a vertical dotted line.}
 \label{fig:solarmodel}
 \end{figure}

Figure~\ref{fig:solarmodel} displays for the points in a standard
solar model
the expected uncertainties in interpolating the radiative opacity
within
opacity tables that have our preferred grid spacings, namely,
$\Delta
\log T = 0.025$ and $\Delta r = 0.125$. In computing the expected
uncertainties shown in
Figure~\ref{fig:solarmodel}, we created simulated OPAL tables at
shifted points in all three
variables: $T$, $r$, and $Z$. We then used the simulated tables to
calculate the opacity at
points (defined by $T$, $r$, and $Z$) that correspond to points in
the standard solar model.
The opacity obtained from shifted simulated tables was compared with
the opacity obtained
from unshifted simulated opacity tables. The differences,
$\Delta \kappa/\kappa$,
between shifted and unshifted simulated opacities are plotted
in Figure~\ref{fig:solarmodel} as a function of the radial
position, $R/R_\odot$, in the solar model and also as a function
of  the
corresponding values of either $\log T$ or $\log r$.  The dotted
vertical line indicates the
location of the base of the convective zone.

We conclude from Figure~\ref{fig:opacityshift} and
Figure~\ref{fig:solarmodel} that opacity
 tables with grid sizes of $\Delta \log T = 0.025$ and $\Delta
r = 0.125$ are probably accurate enough to permit a precise
calculation of the depth of the solar CZ using existing stellar
evolution codes. For the dense grid sizes considered here, the
interpolations within the opacity tables should not cause errors
that prevent an accurate calculation of the depth of the solar
convective zone. However, the absolute value of the tabulated
radiative opacities could still introduce significant
uncertainties.

\section{SUMMARY AND DISCUSSION}
\label{sec:summary}

The primary goal of this paper is to determine how accurately the
radiative opacity near the base of the convective zone must be
known in order to use measurements of the CZ depth to draw
conclusions about other solar parameters. There are two separate
but related issues with respect to the accuracy of the radiative
opacity, namely, the accuracy with which the tabulated values in
opacity tables are calculated and the accuracy with which the
opacity can be interpolated within tables of a specified grid
size. We first summarize our conclusions regarding the accuracy of
tabulated opacity values and then we summarize our results with
respect to the accuracy of interpolations within the standard OPAL
opacity tables. The helioseismological implications of the opacity
changes considered in this paper will be discussed in Bahcall,
Basu, Pinsonneault, and Serenelli, (2004, in preparation).

We show in \S~\ref{sec:dependence} that the logarithmic derivative
of the convective zone depth with respect to the logarithm of the
opacity satisfies $\partial \ln R_{\rm CZ}/\partial \ln \kappa
\approx -0.095$. We conclude from this relation that the radiative
opacity must be known to an accuracy of 1\% in order to calculate
in a solar model the depth of the CZ to the accuracy, 0.14\%, with
which the depth is measured by helioseismology. On the other hand,
if one accepts the recent measurements of heavy element
abundances, then the OPAL opacities must be increased by about
21\% in order to reconcile the calculated solar model depth of the
CZ and the measured depth of the CZ. This change of 21\% could
conceivably arise from a combination of errors in the tabulated
values of the opacity and interpolation errors, which are
discussed below. However, as we shall see, the total change of
21\% is too large to be ascribed solely to errors in
interpolation.

It would be very instructive to have a comprehensive study of the
absolute accuracy of state-of-the-art radiative opacity
calculations. A detailed comparison of the calculated opacity near
the base of the convective zone obtained by the Opacity Project
(Seaton, Yan, Mihalis, \& Pradhan 1994)~\nocite{opacityproject}
with the results of the OPAL project (Iglesias \& Rogers 1996)
would be very informative. The interested reader is referred to
the informative and insightful comparison by Neuforge-Verheecke et
al. (2001) of the Los Alamos LEDCOP opacities and the OPAL
opacities. The largest differences are found near the base of the
convective zone, with the OPAL opacities being as much as 6\%
larger than the LEDCOP opacities in this region. As part of a
comprehensive discussion of factors that affect the accuracy of
solar models, Boothroyd \& Sackmann (2003)~\nocite{sackmann03}
have investigated ways that the opacities can affect
helioseismological parameters.

We show in \S~\ref{sec:composition} that the radiative opacity
near the base of the convective zone depends sensitively upon the
assumed chemical composition (see especially
equation~\ref{eq:kappaoncomposition} and
equation~\ref{eq:compositionderivatives}). If one wanted to calculate
the depth to an accuracy of $0.6$\%, then one would need to know
the heavy element mass fraction, $Z$, to an accuracy of 1\%. This
precision is far beyond the current state-of-the-art accuracy in
the determination of the heavy element abundance.

The entire difference between the measured depth of the solar
convective zone (equation~\ref{eq:rczmeasured}) and the value
calculated  using a solar model with the recent low determinations
of the heavy element abundances (equation~\ref{eq:rcznewzoverx}) could
be explained by the present uncertainty, $\sim 15$\%, in the ratio
of Z/X (see Bahcall \& Pinsonneault 2004)\nocite{BP04}.  Of
course, the changes in opacity caused by changing $Z/X$ are not
limited to any particular region. Changing the assumed surface
value of $Z/X$  affects the composition and hence the opacity
throughout the solar model.

We have approximated in this paper the dependence of the opacity
upon composition by the dependence upon just two variables,  the
mass fractions $X$ and $Z$.  In reality, the situation is more
complex. Different chemical elements contribute differently to the
stellar opacity. For example, Bahcall, Pinsonneault, and Basu
(2001) found that the depth of the convective zone was most
sensitive to the abundances of the lighter metals, which are
significant opacity sources at $2 \times 10^6$K, while the heavier
metals were much more important for the core structure and the
estimated initial solar helium abundance.  However, we are not yet
at a level of precision that we can specify well the
opacity-weighted uncertainties of the different heavy elements.
This is a refinement that will have to await further progress in
determining the different heavy element abundances and more
extensive opacity calculations.

We compare in \S~\ref{sec:comparison} the radiative opacity values
obtained with two different interpolation routines from the
standard OPAL opacity tables. We find that the difference in
interpolated values of the radiative opacity can be as large as
4\% near the base of the convective zone. We also tested in
\S~\ref{sec:shifted} the accuracy with which interpolations can be
performed within simulated opacity tables of different grid sizes.
We find that errors of the order of 3\% may be expected from
tables with the grid spacings of the existing OPAL tables.
However, we show that the interpolation uncertainties could be
reduced to the level of 1\% or below by using a denser grid with
$\Delta \log T = 0.025$, $\Delta \log r = 0.125$, and with $Z$
ranging from $Z = 0.0100$ to
 $Z = 0.0225$ with $\Delta Z = 0.0025$.

For completeness, we report in the Appendix on the calculated
depth of the CZ that was found using  four different equations of
state. In agreement with other authors, we find that the choice of
equation of state affects the calculated depth of the CZ by only
about $\pm 0.1$ \%. We also show in the Appendix that current
uncertainties in nuclear reaction rates also affect the calculated
depth of the convective zone at the level of 0.1\%.

\acknowledgments

JNB and AMS are supported in part by NSF grant PHY0070928 to the
Institute for Advanced Study. We are grateful to M. Asplund for
valuable discussions. AMS acknowledges A. Weiss for providing the
Garching stellar evolution
  code used in  this paper and for useful comments, and H. Schlattl for
  help in learning to use the code.  AMS is supported in part
  by the W. M. Keck Foundation grant to the Institute for Advanced
  Study.

\appendix
\section{IS THE EQUATION OF STATE THE CULPRIT?}
\label{sec:appendix} In order to  estimate the influence of the
equation of state (EOS) on the calculated depth of the convection
zone, we have evolved a series of solar models using different
equations of state. In addition to the OPAL 1996 EOS, we have used
an updated version of the OPAL EOS (OPAL 2001; Rogers
2001)\nocite{OPAL2001}, the MHD EOS (Mihalas, D\"appen, \& Hummer
1988)\nocite{MHDEOS} and the IRWIN EOS.(Cassisi, Salaris, \& Irwin
2003)\nocite{IRWINEOS}

Table~\ref{table:eos} summarizes our results. The variation in the
calculated depth of the convective zone due to varying the assumed
equation of state is
\begin{equation}
\frac{\Delta R_{\rm CZ}}{R_{\rm CZ}} ~\simeq~ 0.001 \, .
\label{eq:eosvariation}
\end{equation}
This variation is similar to the quoted uncertainty in the
measured depth of the convective zone (see
equation~[\ref{eq:rczmeasured}]), but much smaller than the change in
the calculated $R_{\rm CZ}$ required to obtain consistency with
the new, lower heavy element abundances (see
equation~[\ref{eq:rcznewzoverx}]). Similar results have  been found
previously by other authors (Schlattl 2002;\nocite{schlattl02}
Basu, D\"appen, \& Nayfonov 1999)\nocite{schlattl02,basu99},  who
used, however, the larger value of $\left( Z/X \right)_\odot$=
0.0245.

We therefore conclude that the pressure-temperature-density
relationship from the equation of state is not a major component
of the overall error budget for the depth of the solar surface
convection zone. However, the ionization balance of heavy elements
as a function of the physical conditions can have a significant
impact on the opacities; in this indirect sense, the equation of
state will have an impact on the problem.

\vskip 1cm

\begin{table}[!t]
\begin{center}
\caption{Depth of the convection zone in solar radius for
different EOS. \label{table:eos}}
\begin{tabular}{lc}
\hline\hline EOS & $R_{\rm bce}/{\rm R_\odot}$ \\
\hline
OPAL 1996 & 0.7155 \\
OPAL 2001 & 0.7157 \\
MHD & 0.7164 \\
IRWIN & 0.7146 \\
\hline
\end{tabular}
\end{center}
\end{table}

\section{How much effect do nuclear reactions have on the
calculated depth of the convective zone?}
\label{sec:appendixreactions}

For completeness, we record here the small effect that the rates
of nuclear reactions have on the calculated depth of the solar
convective zone.  In Table~1 of Bahcall and Pinsonneault (2004),
the neutrino fluxes are listed for two models, BP00 and New
Nuclear, that differ only in the adopted nuclear reactions. The
New Nuclear model was computed using the best-estimate nuclear
rates as of the end of 2003, while the model BP00 was computed
using the best rates available in 2000.  The computed depths for
the convective zone are $0.7140 R_\odot$ (for BP00) and $0.7147
R_\odot$ (for New Nuclear). Thus, the current uncertainties in the
nuclear reaction rates affect the calculated depth of the solar
convective zone at the level of $0.1$\%.

\section{The conversion of carbon and oxygen to nitrogen during
CNO burning} \label{sec:appendixcnoburning}

During the course of CNO burning, nearly all of $^{12}$C and a
fraction of $^{16}$O are converted to $^{14}$N (for the earliest
discussion of this process, see Section~II.C.2 of Bahcall and
Ulrich 1988\nocite{COburningtoN} and also Section~III.A of Bahcall and Pinsonneault
1992). This process increases slightly (decreases slightly) the
heavy element (hydrogen) mass fraction since, for example, two
protons are added to $^{12}$C to make $^{14}$N.

Unfortunately, the enhancement of $^{14}$N at the expenses of
hydrogen cannot be exactly taken into account with the existing
OPAL opacity tables. The existing tables do not allow the
selective enhancement of nitrogen.

We have therefore evolved two different solar models with two
different treatments of the $^{14}$N enhancement. In the first
model, the enhancement is taken into account and absorbed into the
total heavy element abundance, Z. This treatment correctly
accounts for the increase in Z and the decrease of X when
calculating the opacities but, incorrectly, spreads the increased
heavy element abundance among all of the metals according to their
initial relative abundances. Thus, the solar interior opacity is
slightly overestimated. In the second model, we completely ignored
the increase in Z due to the conversion of carbon and oxygen into
nitrogen when  computing the opacities. In this  case, the solar
interior opacity is slightly underestimated.

Fortunately, the fractional difference is only 0.1\% for the
computed depth of the solar convective zone obtained with these
two different approximations.

\end{document}